# Characterizing primary atomization of cryogenic LOX/Nitrogen and LOX/Helium sprays by visualizations coupled to Phase Doppler Interferometry

N. Fdida [a], Y. Mauriot [b], L. Vingert [c], A. Ristori [d] and M. Théron [e]

**Author names and affiliations:**

[a] nicolas.fdida@onera.fr, ONERA – The French Aerospace Lab, Chemin de la Hunière, BP 80100, 91123 Palaiseau, France

[b] yves.mauriot@onera.fr, ONERA – The French Aerospace Lab, Chemin de la Hunière, BP 80100, 91123 Palaiseau, France

[c] lucien.vingert@onera.fr, ONERA – The French Aerospace Lab, Chemin de la Hunière, BP 80100, 91123 Palaiseau, France

[d] arnaud.ristori@onera.fr, ONERA – The French Aerospace Lab, Chemin de la Hunière, BP 80100, 91123 Palaiseau, France

[e] marie.theron@cnes.fr, CNES – Launcher Directorate, 52 Rue Jacques Hillairet, 75012 Paris, France

**Corresponding author:** NicolasFdida, ONERA

## Abstract

There is a need for experimental data in conditions representative injection in rocket engines to validate or initiate droplet formation models used in numerical simulations. A new cryogenic vessel was built upon the MASCOTTE test bench to study the atomization of a single oxygen liquid jet, under non-reactive conditions, with simultaneous optical diagnostics. A test plan was built to explore the fiber-type regime occurring in liquid rocket injection systems, with a fixed Reynolds number and a large range of Weber number and momentum flux ratio, compared to existing studies. High-speed images are used to describe qualitatively the fiber-type regime and to visualize were droplets are present, in order to prepare the drop-size measurements. A Phase Doppler Interferometer is used to measure the size and velocity of droplets produced by atomization of a liquid oxygen jet by a co-flowing gas. Droplet size and velocity measurements were performed with a PDI close to the nozzle exit in order to provide data on droplets produced by the primary atomization process, which can be useful for numerical simulations initialisation. The radial evolutions of the axial velocity and of the drop size distribution show similar trends as observed in the literature. The axial velocity is investigated for different operating conditions with helium or nitrogen as atomizing gas, showing an increase on the side of the spray. The radial evolution of the droplet size shows a translation of the drop size distribution on the edge of the spray towards the smaller sizes, indicating that the biggest liquid elements stay close to the LOX jet.

**Keywords:** Liquid Propulsion Rocket Engines, Droplet sizing, Cryogenic flows, two-phase flows, Shadowgraphy, Phase Doppler Interferometry

**Abbreviations:**

BVF: Flow Visualization Box

CNES : Centre National d'Etudes Spatiales

LOX: Liquid Oxygen

MASCOTTE : Montage Autonome Simplifié pour la Cryocombustion dans l'Oxygène et Toutes Techniques Expérimentales

ONERA : Office National de Recherches Aérospatiales

pdf : Probability Density Function

PDI : Phase Doppler Interferometer

**1. Introduction**

In liquid propellant rocket engine combustion, physical processes such as atomization, mixing or vaporization of propellants are complex and can interact with one another. Thermo-acoustic instabilities can develop during the transient operating states of the engine and lead to its deterioration. Atomization is a dominant process that drives the behaviour of such a cryogenic flame, particularly when the propellants are injected in subcritical conditions, investigated by several research teams ([1], [2], [3] and [4]). Atomization therefore needs to be studied in order to validate CFD codes which will be used to predict interactions between acoustics and atomization, and possibly the occurrence of these high instabilities. ONERA and CNES have a common interest to improve the knowledge on cryogenic atomization in liquid propellant rocket engines, in order to build reliable physical models and validation databases for CFD codes [5]. The validation and initialization of recent numerical simulations ([5] and [6]) requires experimental data in the atomization zone, situated close to the injector inner liquid oxygen (LOX) post. Thus we performed here droplet size measurements closer to the nozzle exit than in other previous studies ([3], [7]) to provide data for initialisation of numerical simulations. Fig. 1 shows a qualitative comparison between experiments and numerical results on an instantaneous field obtained at ONERA. An experimental shadowgraph of a LOX/$H_2$ cryogenic jet flame obtained at a chamber pressure of 1 MPa [7] is presented on the left of Fig. 1 and an iso-surface of 95% of the LOX mass fraction obtained by CFD in the same reacting conditions [5] is presented on the right of Fig. 1. In such a two-phase flow, combustion makes measurements and simulations very tough, so atomization is studied first without the presence of the flame in order to predict the spray formation more efficiently.

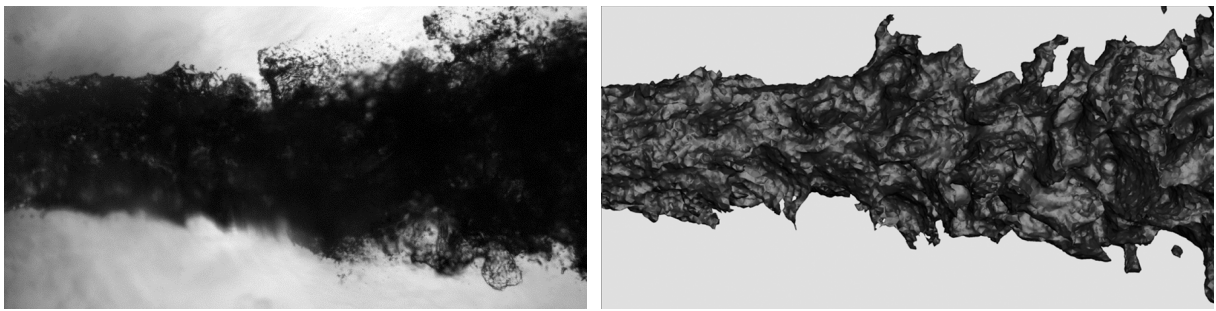

Fig. 1: Comparison between experiments (left) and numerical results (right) performed at ONERA.

A new experimental configuration, presented by Mauriot et al., 2016 [8], is built upon the MASCOTTE test bench. It is dedicated to study the atomization of the liquid oxygen jet with simultaneous optical diagnostics. This cryogenic Flow Visualization Box (BVF) is fully instrumented with pressure and temperature transducers and can be pressurized in a wide range of chamber pressures, flow rates and fluid temperatures under conditions which are inert but nonetheless representative of industrial conditions concerning the fluids injection. The goal is to measure droplet sizes produced by the primary atomization process and, in the future, to link their size with the physical parameters at injection, according to the literature [9], [10]. Completing the work of Farago and Chigier [11], Lasheras and Hopfinger [12] proposed a classification of breakup regimes for a coaxial jet as a function of three parameters: the gaseous Weber number $We_G = \rho_g U_g^2 D_L/\sigma$ (based on the LOX post diameter $D_L$, the gas density $\rho_g$, the gas velocity $U_g$ and the surface tension $\sigma$), the Reynolds number of the liquid $Re_L = U_L D_L/\nu_L$ (based on the LOX post diameter $D_L$, the liquid jet velocity $U_L$ and its viscosity $\nu_L$) and the gas-to-liquid momentum flux ratio $J= \rho_g U_g^2/\rho_L U_L^2$. The typical atomization regime occurring in a rocket engine is defined as the fiber-type injection, i.e. high gaseous Weber number ($We_G > 10^3$) and Reynolds number ($Re_L > 10^4$) and a momentum flux ratio $J > 1$. Compared to previous studies, the operating conditions are similar in terms of Reynolds and Weber numbers, but in this paper the influences of $We_G$ on droplet sizes and velocities are studied whereas Pal et al., 1996 [3] and Porcheron et al., 2002 [13] studied respectively the effect of the $Re_L$ and the injection gas density on the atomization. Most of the studies on coaxial atomization were performed with water and air at atmospheric pressure. Nevertheless, Porcheron et al., 2002 [13] showed that the density of the atomizing gas has a significant influence on the extension of the intact liquid core and thus on the atomization process, comparing the results with three different atomizing gases (nitrogen, helium and argon). To characterize the atomization process in representative conditions, liquid oxygen and two different atomizing gases have been used in our study.

The BVF was designed to operate simultaneously high-speed diagnostics such as Phase Doppler interferometry (PDI) and high-speed visualization. Coupling those complementary diagnostics is very useful to understand the dynamics of atomization considered as a multiscale phenomenon. Thus, imaging provides large as well as close field of views, according to the optical setup, to visualize either the overall spray or focus on the droplet sizes or velocities [7]. However, in the present operating conditions, the use of the PDI is more adapted than imaging to measure small droplet sizes and high velocities close to the injector. High-speed visualizations are used to characterize the dynamics of the atomization process of such a spray and also to localize the PDI measurement volume where droplets are present. The PDI measurement locations are closer to the injector exit than in previous studies, to measure droplets produced by the primary atomization process which can be linked to the spray injection parameters [9], [10].

## 2. Materials and Methods

*2.1 The MASCOTTE Test Bench*

*2.1.1 The BVF Flow Visualization test vessel*

The MASCOTTE test facility was developed at ONERA in order to investigate physical phenomena involved in the combustion of cryogenic propellants [14], such as atomization, ignition, combustion instabilities, thermal transfers at walls, … Among these, atomization remains a key issue, especially in subcritical pressure conditions, i.e., when LOX is injected in the combustion chamber below the oxygen critical pressure $P_c (O_2) = 5.04$ MPa and critical temperature $T_c (O_2) = 154.6$ K. To decouple the atomization process from combustion, a new cryogenic test vessel, called BVF, was built to study a LOX jet atomized with cold gaseous $N_2$ or He. The BVF is designed with large inner dimensions (381 mm in diameter by 995 mm in height) to limit the spray interaction with the walls and the windows. Moreover, the internal part of the windows is protected from droplets impacts by a gaseous film on the windows. An external heating device protects the windows from icing, as seen on Fig. 2. The BVF is fully instrumented in pressure and temperature transducers adapted to cryogenic environment. It is compatible with laser measurements close to the injector exit, with a pressurization up to *Pc* < 3 MPa and is able to reach fiber-type injection conditions, representative of a liquid rocket engine injection device. The four quartz windows have the same dimensions (optical aperture of 153 mm in height and 103 mm in width) and are located at the same height; three of them are placed at 90° of each other, the fourth one being placed at an angle optimized for Phase Doppler droplet sizing (145° between emitter and receiver). The total duration of a typical test sequence is about 130 seconds whereas the stationary phase of the flow is about 40 seconds. This 40 seconds duration results from a compromise between a reasonable time for the PDI to collect a sufficient number of droplets and the need to characterize an operating point (by 5 to 10 PDI measurement locations), without running out of fluids.

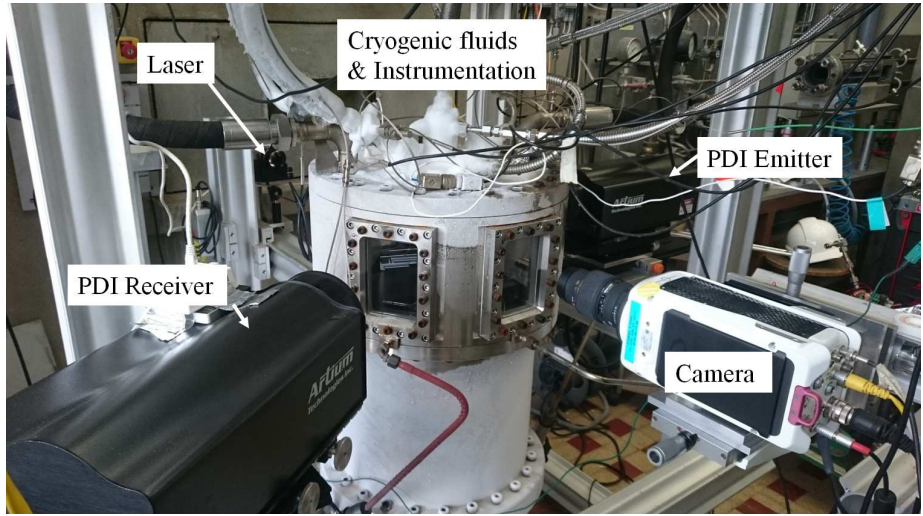

Fig. 2: Optical setup around the cryogenic test chamber.

*2.1.2 Injection configuration*

The cryogenic fluid lines are regulated in temperature from the tank to the injection head of the vessel. The atomizing gas injection temperature is maintained as close as possible to the LOX temperature (difference lower than 10 K) in order to minimize the influence of vaporization on the droplet sizes, which could introduce a bias on the size measurements. For one test case, the temperature deviation is about 1 degree during the stationary phase of the flow. In liquid propellant rocket engines, running with LOx/H2, the use of shear-coaxial type elements in the injector assembly is common. The single coaxial injector used in the BVF produces a central low-velocity LOX jet surrounded by an outer high velocity co-flowing gas, as illustrated by Fig. 3. The LOX jet issues from a cylindrical orifice characterized by its internal diameter $D_L$ and the gas stream exits from a coaxial annular slit defined by its internal diameter $D_G$. The mixing between both fluids occurs outside from the injector with no velocity swirl component in the gas stream. Thus, the gas and liquid flows enter in interaction with each other, leaving their respective nozzle with the same direction. Each stream is characterized by its own average velocity noted $U_L$ and $U_g$, for the liquid and the gas, respectively, which are evaluated by pressure transducers placed in the injection head. The coaxial injector is visible inside the field of view, in order to see the LOX jet from the exit of the injector to a distance of it equal to several times its diameter.

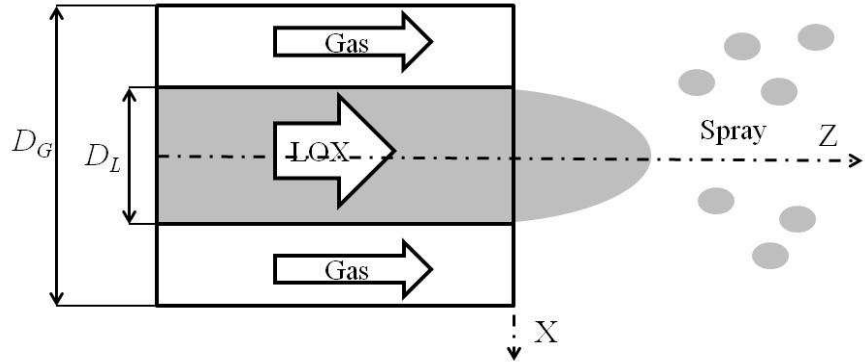

Fig. 3: Schematic view of the coaxial injection configuration

The test plan explores the fiber breakup regime with parameters driving such an atomization process: the gas velocity, the injector geometry and the atomizing gas density which take part of the linear stability analysis of a liquid/gas shear interface theory [12]. The operating points shown on the ($We_G$, $Re_L$) diagram on Fig. 4 are distributed along a fixed Reynolds line, with variations of the Weber number as well as the $J$ number. On this diagram, symbols represent the operating conditions investigated during two experimental campaigns performed in 2015 and 2016. The generic operating conditions are called xN or xH, respectively for Nitrogen and Helium as atomizing gas, where "x" is an arbitrary case identifier. 8Nter and 8H were chosen so as to have $J$ number values of the as close as possible. This graph permits to compare the operating conditions with similar existing studies, such as the one of Pal et al., 1996 [3], where $Re_L$ varies from 10000 to 97000, with a constant $We_G$ =4300, but in a water/N$_2$ spray. Indeed most of the studies on non reacting flows have been performed with water/air or water/N$_2$. Lines of constant $J$ are shown as straight lines in logarithmic coordinates, for constant values of $D_L$ and $U_L$, according to the relation $Re_L = (We_G/J)^{1/2}(D_L \sigma/\rho_L U_L^2)$, the dashed line corresponding to $J$=1 in the water/air experiment [12] and the solid ones to a liquid oxygen/inert gas mixing. In the study of Porcheron et al., 2002 [13], the operating conditions are close to our configuration, with $Re_L$ =68200 and $We_G$ varying from 12000 to 14000 and $J$ fixed at 6. In our study, the momentum flux is varying in the range 1< $J$ < 12 and the dimensionless number domain is within the fiber-type regime, with 2000 < $We_G$ < 12000 and a fixed Reynolds number $Re_L \approx$ 65000. The $We_G$ varies from 2000 to 11000, under the effect of the gas injection velocity $U_g$ and the gas density $\rho_g$. These dimensionless numbers are calculated during the stationary phase of the flow with the pressure and temperature transducers that equip the BVF. Nine typical instantaneous images are also shown on this diagram.

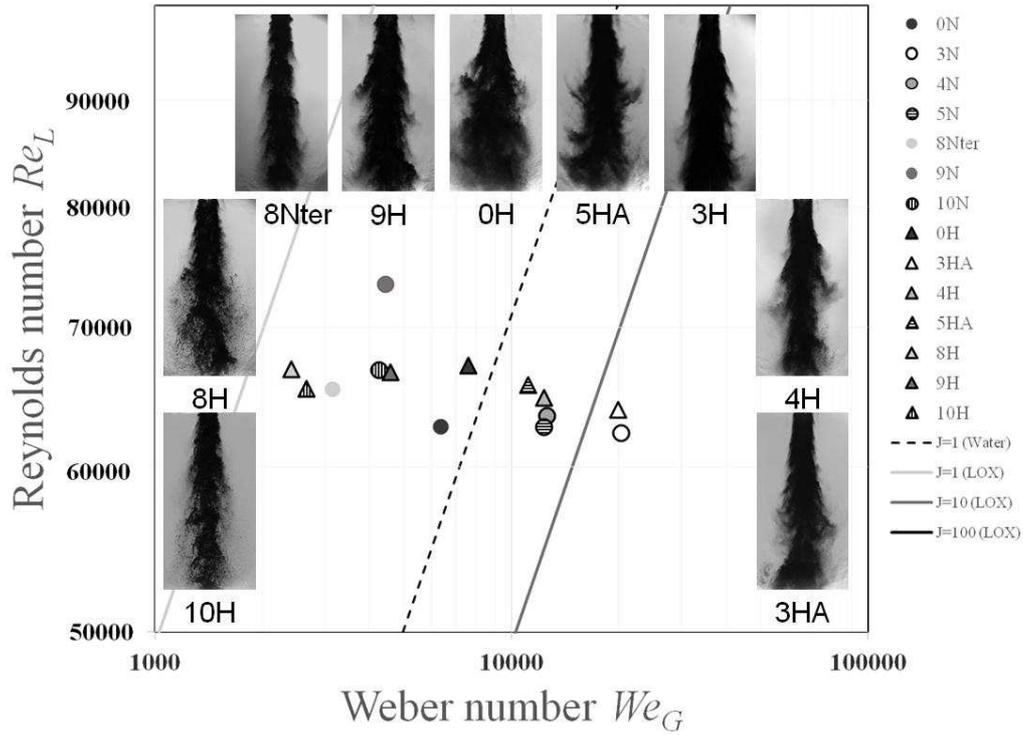

Fig. 4: Injection parameters of the fiber-type regime in ($We_G$, $Re_L$) diagram.

The test plan was built to explore the LOX jet atomization in a fiber-type break-up regime, which is characterized by the creation of very thin and short liquid fibers created from the continuous liquid jet exiting from the nozzle. These fibers are rapidly peeled off the jet, stretched by the differential velocity between the liquid jet and the outer gas stream, from which the Weber number depends. The characteristic break-up time is very short, which means a rapid atomization of the majority of the liquid. The droplets are produced in very small sizes, several orders of magnitude smaller than the injection diameter. High speed imaging is then necessary to capture the spray dynamics. A video recorded at 16 kHz illustrating the dynamics of the fiber-type atomization process in such injection conditions is linked to this paper and the still image presented on the left of Fig. 5, extracted from the linked video, illustrates the whole process by an overall view of the spray. As written previously, the atomization process is initiated by a longitudinal wave, which is evidenced by the right pictures of Fig. 5 whereas the left picture illustrates the accomplished atomization process. Some droplets resulting from this process may then be deformed and fragmented by secondary atomization as explained by Guildenbecher et al. [15].

An illustration of the atomization initialization process is shown on the right of Fig. 5 with a set of successive images recorded at 16 kHz for the injection condition 0H. The injector exit is placed on the left edge of each image; the flow is going from the left to the right and time passes from top to bottom. Except for the last still image, only the upper half of the jet is showed. As explained by Marmottant and Villermaux [9], a longitudinal primary instability, created at the injector exit by shear instability, develops an undulation on the jet surface. This undulation is growing in amplitude, with crests accelerated by the high velocity coflowing gas and transformed into fibers by a transverse instability process. Due to this stretching, the liquid jet is highly modified: those fibers are then rapidly elongated into ligaments which collapse into droplets further downstream [16].

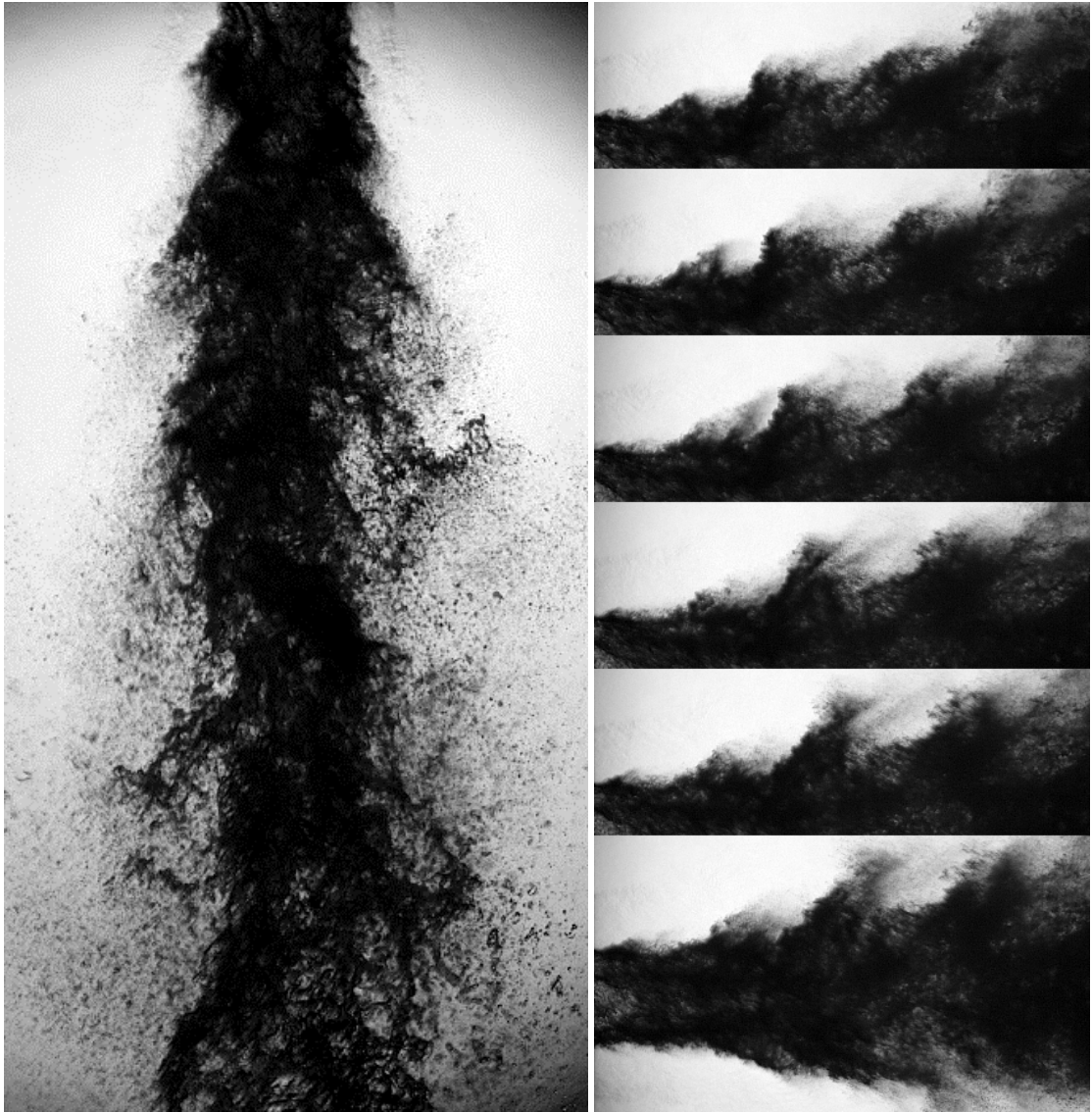

Fig. 5: Illustration of the primary instability (right) and of the achieved atomization process (left) with instantaneous successive images recorded at 16 kHz for 0H case (right) and 8H case (left).

*2.2 Optical Setup*

A PDI is used simultaneously with a high-speed camera thanks to the particular four windows arrangement on the BVF.

*2.2.1 High-speed shadowgraphy*

A high-speed camera shadowgraphy setup is used to visualize the droplets location in the spray in order to detect the region of interest for the spray exploration with the PDI. The spray is enlightened in a backlight configuration by a laser diode CAVILUX Smart 400W which emits a red incoherent light pulse at 640 ±10 nm. This source is very compact and stable in space and time, enlightening the LOX jet through an optical fiber. The pulse duration can be set in a large range of time periods, from 0.02 µs to 10 µs, with 0.01 µs steps. In our application where droplets velocities are very fast, we fixed the pulse duration to 0.04 µs to freeze the droplets on the images and avoid any blurring effect. The repetition rate can be set from 25 Hz to 10 kHz continuously and can be increased to 50 kHz during limited periods. A Fourier lens of 50 mm focal length is placed at the fiber exit, as well as a diffuser to create a large illumination area to see the spray in a large field of view, as shown on Fig. 5.

Images of the spray were recorded with a Phantom v711 high-speed camera from Vision Research. This camera is composed of a 12-bit CMOS sensor of 1280 x 800 square pixels and of 20 µm side length. The resolution of the sensor is directly related to the frequency rate. The sensor resolution was set at 800 x 400 px² at 25 kHz or 1024 x 512 px² at 16 kHz. The camera is equipped with a Sigma lens of 105 mm focal length and a narrow band pass filter centred on the red illumination wavelength of the light source. Hence green and blue beams coming from the PDI are not recorded by the camera.

*2.2.2 Phase Doppler Interferometer (PDI)*

The Phase Doppler Interferometer, from ARTIUM Inc, is a particle counter which was used to measure the oxygen droplet size and velocity distribution in the LOX spray, under steady operating conditions. A solid state laser system delivers green ($\lambda_g$ = 532 nm) and blue beams ($\lambda_b$ = 491.5 nm). The focal length of the emitter lenses was set to 350 mm and the one of the receiver was set to 500 mm, constituting a setup adapted for such a dense spray. The off-axis angle of the receiver was set to 35° as recommended by the manufacturer, for the expected droplet size range. Indeed, it results from a compromise between 40°, which is more favorable to the measurement of droplets of diameter between 1 and 10 microns, and 30° which is more suited to the measurement of larger droplets, according to the manufacturer's recommendations. The angle of collection, set at 35 ± 1°, can induce a systematic error estimated at 2% on the measured diameters in this setup. More details on the basics of this instrument are presented by the designers of the instrument [17]. The PDI performs a sphericity check for each droplet passing inside the measurement volume thanks to a four channel detector [18]. The droplet sphericity is evaluated with an estimation of its diameter in two perpendicular planes. Droplets are rejected by the PDI when these diameter estimations are different. The PDI was set with a 50 µm slit aperture on the receiver to collect a thin part of the light coming from the measurement volume. Indeed, in order to avoid multiple droplet signals we chose a short focal length at emission together with a small aperture on the receiver, which resulted in a short measurement volume [19].

The estimated refractive index *n* of the LOX droplets indicated to the PDI is fixed to *n*=1.21, according to the work of Johns and Wilheim [20], in order to ensure an uncertainty of ±2% on the diameter size measurement in the temperature range from injection to boiling temperature in the test chamber thermodynamic conditions. Based upon the presented optical setup, the PDI could measure droplet diameters within the range 1 µm < *D* < 108 µm. The PDI provides two components of the droplet velocity (horizontal *Vx* and vertical velocity *Vz*, respectively through blue and green channels). The largest velocity range in our acquisitions was -20 m/s < *Vx* < 20 m/s and -75 m/s < *Vz* < 200 m/s for horizontal and vertical velocity, respectively. The directions of the measured velocities are defined on Fig. 6.

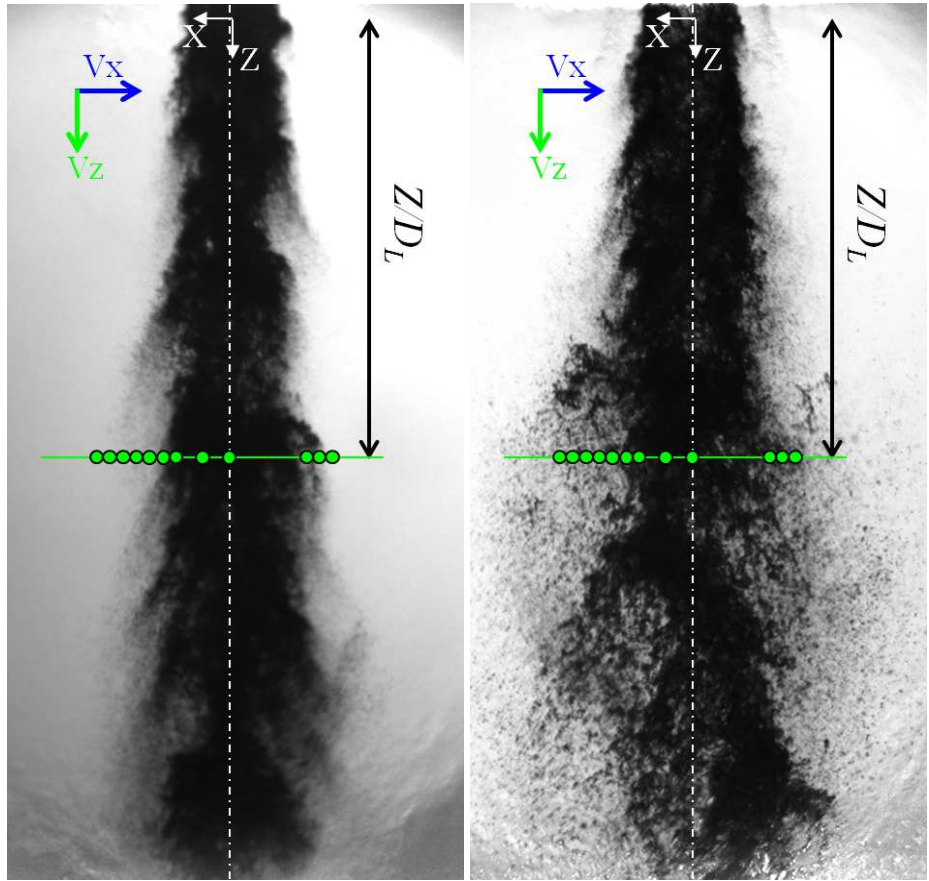

a) LOX/N$_2$ injection, 8Nter          b) LOX/He injection, 8H

Fig. 6: Location of PDI measurements from the injector inner post; $V_x$, $V_z$ are velocities measured by PDI.

*2.2.3 Measurement locations*

By nature, PDI measurements can only be performed in regions where droplets are present. Images recorded by the high-speed camera are helpful for PDI to identify these regions. An example of instantaneous shadowgraphs is presented in the background of Fig. 6, superimposed with the PDI measurement locations, illustrated by green dots. On the instantaneous image of Fig.6.a, there are droplets in the PDI measurement volumes located on the right side but not on the left side. But considering the dynamics of the atomization process illustrated by Fig.5, and as the PDI is a particle counter, droplets may also be collected on the left side, until the extreme left PDI measurement position. Further outside from the spray centerline however, there is not enough droplets passing through the PDI measurement volume to obtain a statistically robust sample of droplets. PDI measurements are performed in a vertical plane Y=0, including the injection axis. In this plane, at a vertical position $Z/D_L$, transverse profiles were made along the *X* axis, with 5 to 10 measurement points radially distributed along the LOX spray, in the two-phase flow area. Diameter and velocity measurements were made at a close distance from the injector exit $Z/D_L$, with $Z/D_L < 5$. The vertical position of the PDI measurement volume is chosen according to the injection parameters as $Z/D_L = 6/(J)^{0.5}$. Indeed, at this position corresponding to the theoretical penetration length, the droplets are supposed to be produced by the primary atomization process which could be linked to the injection parameters, according to [9]. The coordinates of the PDI measurement volume are known, relatively to the injector reference system, thanks to a graduated object that is mechanically fixed to the injector. It ensures an uncertainty in position of ±0.5 mm and ±1 mm, respectively on X and Y axis, corresponding to the PDI optical axis. For each operating condition, a set of 5 to 10 PDI measurement locations is obtained depending if droplets are actually collected at each location.

Droplets measurements are collected during 35-40 s in the stationary phase of the flow. The number of droplets obtained in the sample varies from few ones to several thousands, depending on the PDI validation rate. The validation rate is the percentage of droplets that are accepted by the instrument as being spherical drops. It varied from a few percents to 64% during this campaign, depending on the PDI location in the spray. The lowest validation rates are obtained close to, or inside the liquid jet ($|X/D_L| < 0.5$). The highest validation rate reaches 64% for the measurement point located at the center of the line of points ($|X/D_L| \sim 1$), with 124000 droplets detected in 25 s. To achieve a reasonably accurate estimate of the mean diameter $D_{10}$ of the droplet sample, samples of at least 1500 measurements are necessary (which allows a +/-10% accuracy on $D_{10}$ in the case of a gaussian distribution [21]).

## 3. Results and Discussion

### 3.1 High-speed Visualization

Time-averaged (top) and RMS (bottom) images are shown on Fig. 7 for four operating points with LOX/N$_2$ and LOX/He sprays, increasing $J$ and $We_G$. Time Averaged and RMS images have been calculated on 4000 successive images recorded during the stationary phase of the flow. Increasing the Weber number tends to enlarge the spray plume, as it can be seen on time averaged images between 4H and 3HA cases (the injector geometry is fixed and the helium gas velocity is increasing). The highest fluctuations, in bright blue on RMS images, can be interpreted as corresponding to the area of droplet production whereas the dark blue regions correspond to monophasic areas, that can be only LOX jet or mainly gaseous phase. It implies that the PDI measurement volume has to be preferentially located in those bright blue area to collect droplets. This bright blue zone of main production of droplets is confined in a narrow region surrounding the LOX jet, which confirms the observation of Pal et al. [3]. Moreover, the RMS images show that higher fluctuations are surrounding the LOX core as the Weber number increases. In this area, which is distributed close to the LOX jet for 3HA, a large proportion of tiny droplets is produced, creating sometimes a wall for light to be collected by PDI or imaging. Considering 0H and 8Nter cases, the bright blue lines are larger in the case of 0H case, which could be interpreted as corresponding to a narrower droplet production area in the 8Nter case, corresponding both to a reduced value of the gaseous Weber number and to N$_2$ as the atomizing gas (N$_2$ is denser than helium by a factor 2.5 comparing these cases). The observed effect should be attributed to the combined influences of $We_G$ and $\rho_G$.

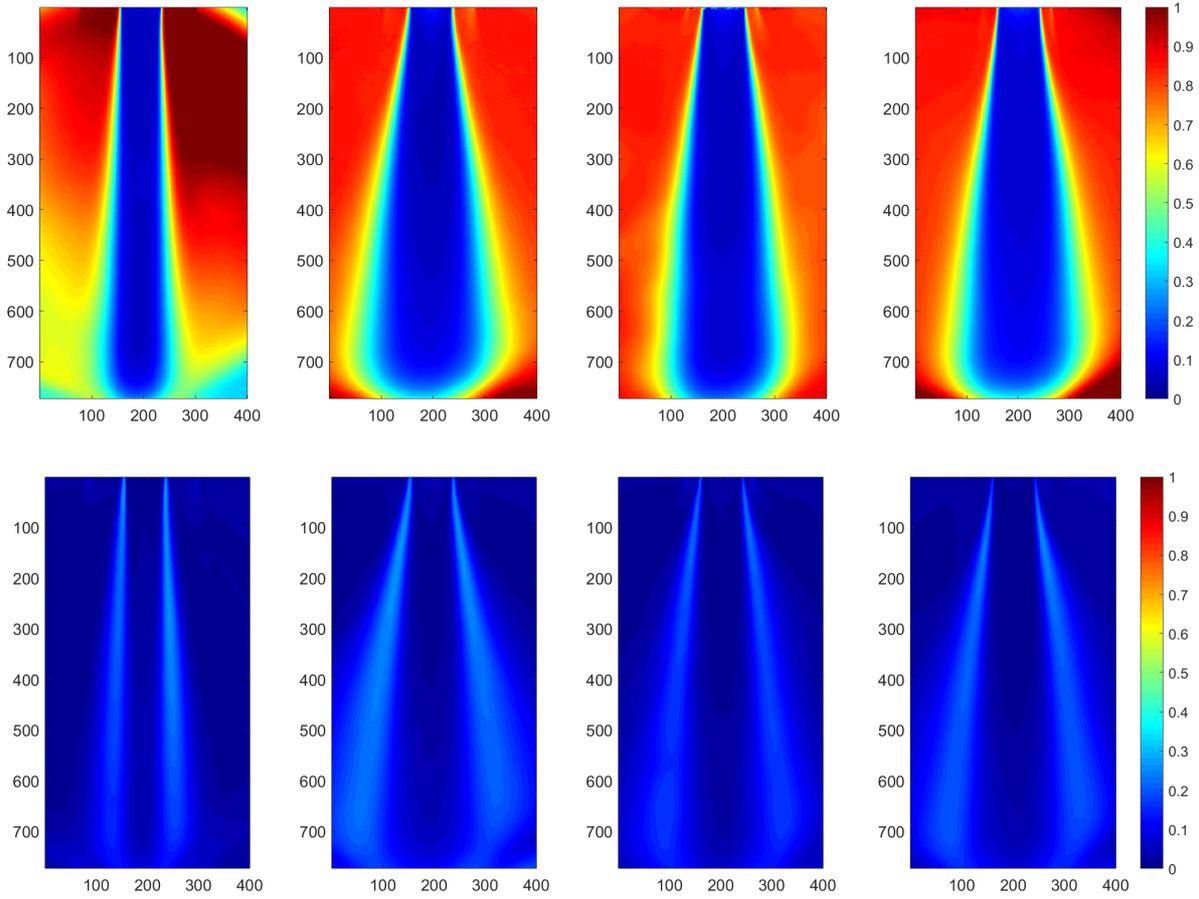

8Nter ($J$=1.9, $We_G$=3.10$^3$)   0H($J$=4.3, $We_G$=7.10$^3$)   4H($J$=7.2, $We_G$=1.10$^4$)   3HA($J$=11.7, $We_G$=2.10$^4$)

Fig. 7: Time averaged (top) and RMS (bottom) images for four operating conditions.

*3.2 Droplet velocities*

The effect of the atomizing gas $N_2$ or He can be studied with PDI measurements on Fig. 8: mean droplet axial velocity $Vz/V_{ref}$ is plotted versus the radial position $X/D_L$ of the PDI measurement volume for three operating points: 8Nter, 8H and 4H. $V_{ref}$ is a reference axial velocity which has been chosen accordingly to the study of Fdida et al., 2016 [7]. The size of each disc representing a PDI measurement is directly proportional to the mean droplet diameter $D_{10}$. We notice that for a defined operating point the more rapid is the axial velocity, the smaller is the mean drop size, indicating that the droplet size is driven by the $We_G$, as proposed by [10]. The velocity radial profiles are not symmetric with respect to the injection axis, showing that the perpendicularity between the orientation of the injector axis and the PDI velocity orientation is barely satisfactory. The increase of the axial velocity with radial distance from the centreline was evidenced by Pal et al. [3] on water/$N_2$ shear coaxial injectors and they also noticed that the velocity reaches a maximum at radial distance $X/D_L > 2$, and then decreases for greater radial distance. Due to a lack of droplets for distance $X/D_L > 1.5$, we did not obtain PDI results so far from the injection axis. This lack of droplets could be explained by the measurement volume location which is closer to the injector outlet ($Z/D_L < 5$) whereas in other studies, the PDI axial location is usually further downstream, at $Z/D_L \geq 10$ for [3] and [13]. Even if the droplet number is low at this position close to the injector outlet, characterizing those droplets is very useful because they result from the primary atomization process, which is of paramount importance for modelling and numerical simulations. Thus information on droplet velocities at such a close position from the injector could be used to initiate numerical simulations.

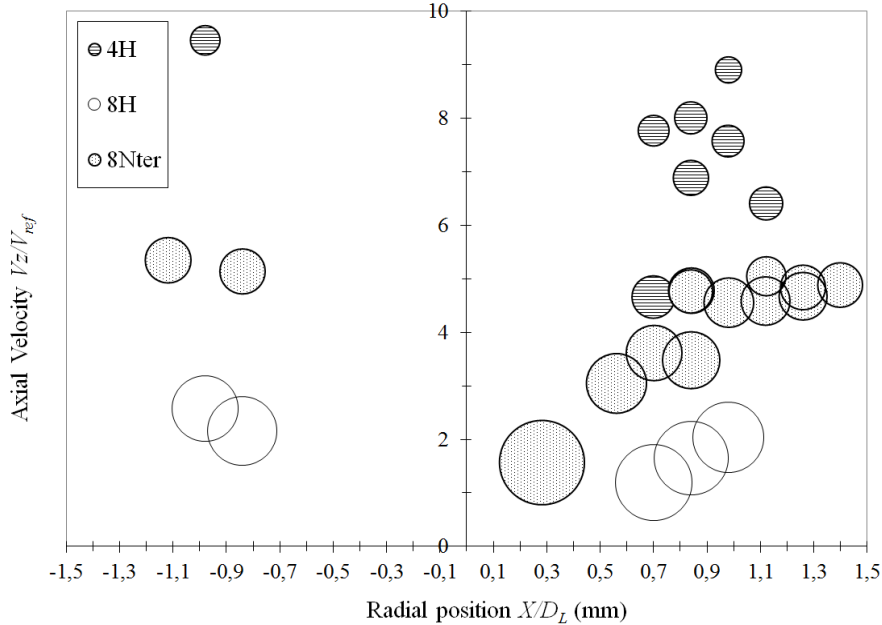

Fig. 8: Overview of the radial velocity profiles measured by PDI in different test cases 8Nter, 8H and 4H. Axial velocity $V_z/V_{ref}$ is plotted versus PDI radial position $X/D_L$ and the size of each disc is directly proportional to the mean droplet diameter $D_{10}$.

*3.3 Drop-size measurements*

From 8Nter case data shown on Fig. 8, the radial evolution of the mean droplet size $D_{10}$ slightly decreases radially until $X/D_L=1$ and then, as the distance from the injector axis increases the trend is less clear. The decrease of mean diameter has also been observed by Pal et al., 1996 [3] on a spray produced by a water/N$_2$ shear coaxial injector and phase Doppler measurements. They also noticed an improvement of primary atomization with the increase of the momentum flux ratio $J$, inducing a decrease of the mean drop size.

The radial evolution of the drop size is also presented on Fig. 9, by the probability density function (pdf) $f_n$ for the operating point 8Nter, with $D_{ref}$ a reference diameter which has been chosen accordingly to the study of Gicquel and Vingert, 2000 [22]. The function $f_n$ gives the probability to have a droplet of diameter $D/D_{ref}$ in the spray, as defined in [23]. The above mentioned trend is clearer with such a diagram: the mean drop size decreases radially as the pdf is translated towards the small sizes. The smallest droplets are encountered at the furthest position from the LOX jet axis, on the edge of the spray for $X/D_L=1.4$. For this position on the edge of the spray, the PDI data rate (droplets collected by second) is relatively low because the droplet density (number by volume) is small, it takes time to acquire data even with a good PDI validation rate (> 60%). Such a high validation rate indicates that most of droplets are spherical, for this position in the spray. As the PDI measurement location comes closer to the LOX jet, the PDI counts more particles by interval of time but the validation rate is lower and usually a validation rate of about 20% is obtained around $X/D_L=1$. On the other side, as the PDI measurement location comes closer to the injection axis, the overall shape of the pdf is translated towards bigger diameters, until $X/D_L = 0.56$. The biggest liquid elements of the drop size stay close to the centre of the liquid jet due to their larger inertia. Droplets are smaller on the edge of the spray due to the effect of aerodynamic forces as well as vaporization, reducing the droplet size.

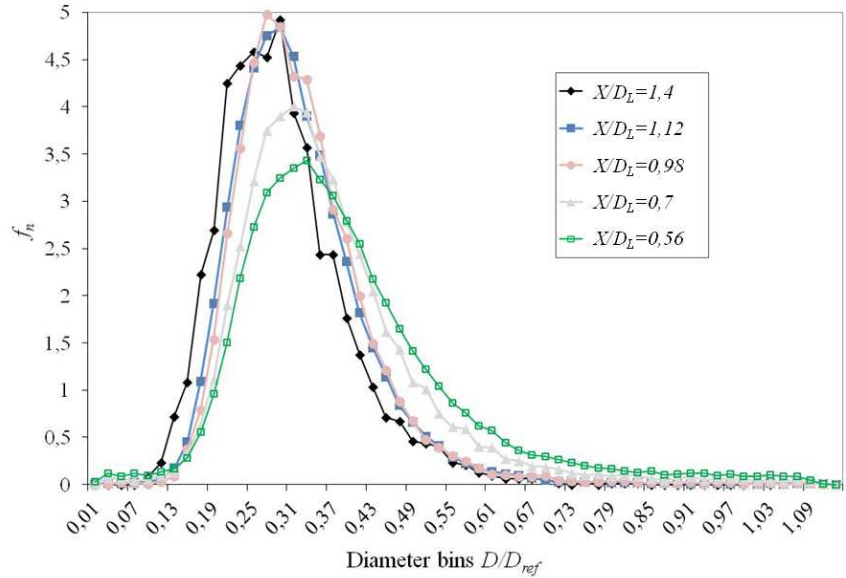

Fig. 9: Radial evolution of the drop size distribution $f_n$ for five radial positions $X/D_L$, with $Z/D_L$=4.5, 8Nter case.

## 4. Conclusion

Understanding the physics of the atomization for a specific injector is important for building theories on the subsequent dynamics of vaporization, mixing and combustion. Those theories can only be validated by experimental data that detail the droplet-size and velocity distributions under various operating conditions, reacting or inert. There is a lack of experimental data in the liquid rocket engine literature because of the harsh environment for optical diagnostics in liquid propellant rocket engines, the safety aspects to consider when working with liquid oxygen and the expensive nature of the experiments. With the support of CNES, this experimental campaign led to drop-size and velocity measurements in various operating conditions in order to describe the spray produced by a shear coaxial injector in non-reactive conditions, representative of a single element of a liquid rocket engine. A test plan was built to explore the fiber-type atomization regime with a fixed liquid Reynolds number and a larger Weber number range compared to previous experiments investigating coaxial injectors. The dynamics of the fiber-type regime is well described by the high-speed images, which were also useful to prepare the drop-size measurements. This experimental test campaign in inert conditions was very fruitful of data in terms of injection conditions and greatly increased the database of MASCOTTE experimental non reacting injection cases. Droplet size and velocity measurements were performed by PDI, closer to the nozzle exit than in other previous studies in order to provide data to validate models and for initialisation of numerical simulations. Indeed, at such a close distance from the injector, even if the droplet number is low, those droplets are important because they are produced by the primary atomization process which is not taken into account in most of numerical simulations in representative conditions. The radial evolution of the axial velocity shows an increase together with the gaseous Weber number. The radial evolution of the drop-size distribution shows an effect of quasi-translation of the pdf towards the small sizes, as the radial position increases, which has already been observed by Pal et al., 1996 [3], but not in such details. This database will help to build atomization models and CFD solvers for predicting initial drop size distributions. Indeed, the locations of the PDI measurement points are close enough to the injector to consider that we investigate the primary atomization process.

Some perspectives of this work are to fit the experimental distributions on mathematical models existing in the literature, such as the log-normal law, the gamma law or the maximum entropy formalism. Moreover, recent results have shown that the droplet mean diameter could to be linked with physical parameters driving this atomization process [24]. Besides, we have experimented reacting cases, measuring the droplet size and velocity, with LOX/$H_2$ or LOX/$CH_4$ jet flames, to address the influence of the flame on the drop size produced by a shear coaxial injector, which will be the subject of a future publication.

**Acknowledgments**

A part of this work has been co-funded by the Centre National d'Etudes Spatiales (CNES) and by ONERA, in the framework of a common interest program dedicated to the study of high frequency instabilities. The authors also thank Mr Carru, Vannier and Paux for their assistance in conducting the experiments.

**Nomenclature**

| | | |
|---|---|---|
| $D_G$ | Injector diameter (gas side) | mm |
| $D_L$ | Injector diameter (LOX side) | mm |
| $D$ | Droplet diameter | μm |
| $D_{ref}$ | Reference diameter | μm |
| $f_n$ | Numerical probability density function | Dimensionless |
| $J$ | Momentum flux ratio | Dimensionless |
| n | Refractive index of the medium | Dimensionless |
| $P_{ch}$ | Pressure in the combustion chamber | MPa |
| $P_c(O_2)$ | Critical pressure of oxygen | MPa |
| $Re_L$ | Reynolds number based on LOX post diameter $D_L$ | Dimensionless |
| $T_c(O_2)$ | Critical temperature of oxygen | K |
| U | Injection velocity of the fluid | m/s |
| $V_{ref}$ | Reference axial velocity | m/s |
| $V_x$ | Radial velocity | m/s |
| $V_z$ | Axial velocity | m/s |
| $We_G$ | Gaseous Weber number | Dimensionless |
| $\lambda$ | Wavelength | nm |
| $\nu$ | Fluid kinematic viscosity | m²/s |
| $\rho$ | Fluid density | kg/m³ |
| $\sigma$ | Surface tension of the fluid | N/m |